\documentclass[aps,prl,twocolumn,showpacs,preprintnumbers,amsmath,superscriptaddress,amssymb,floats,nofootinbib]{revtex4}

\usepackage{graphicx}
\usepackage{dcolumn}
\begin{document}

\def\Journal#1#2#3#4{{#1} {\bf #2}, #3 (#4)}
\def\NCA{\rm Nuovo Cimento}
\def\NIM{\rm Nucl. Instrum. Methods}
\def\NIMA{{\rm Nucl. Instrum. Methods} A}
\def\NPA{{\rm Nucl. Phys.} A}
\def\NPB{{\rm Nucl. Phys.} B}
\def\PLB{{\rm Phys. Lett.}  B}
\def\PRL{\rm Phys. Rev. Lett.}
\def\PRD{{\rm Phys. Rev.} D}
\def\PRC{{\rm Phys. Rev.} C}
\def\ZPC{{\rm Z. Phys.} C}
\def\JPG{{\rm J. Phys.} G}

\title{Cross Section Measurement of Charged Pion Photoproduction\\
 from Hydrogen and Deuterium}
\author{L.~Y.~Zhu}
\affiliation{Massachusetts Institute of Technology, Cambridge, MA
02139, USA}
\author{J.~Arrington}
\affiliation{Argonne National Laboratory, Argonne, IL 60439, USA}
\author{T.~Averett}
\affiliation{College of William and Mary, Williamsburg, VA 23185, USA}
\affiliation{Thomas Jefferson National Accelerator Facility, Newport
News, VA 23606, USA}
\author{E.~Beise}
\affiliation{University of Maryland, College Park, MD 20742, USA}
\author{J.~Calarco}
\affiliation{University of New Hampshire, Durham, NH 03824, USA}
\author{T.~Chang}
\affiliation{University of Illinois, Urbana, IL 61801, USA}
\author{J.~P.~Chen}
\affiliation{Thomas Jefferson National Accelerator Facility, Newport
News, VA 23606, USA}
\author{E.~Chudakov}
\affiliation{Thomas Jefferson National Accelerator Facility, Newport
News, VA 23606, USA}
\author{M.~Coman}
\affiliation{Florida International University, Miami, FL 33199, USA}
\author{B.~Clasie}
\affiliation{Massachusetts Institute of Technology, Cambridge, MA
02139, USA}
\author{C.~Crawford}
\affiliation{Massachusetts Institute of Technology, Cambridge, MA
02139, USA}
\author{S.~Dieterich}
\affiliation{Rutgers University, New Brunswick, NJ 08903, USA}
\author{F.~Dohrmann}
\affiliation{Argonne National Laboratory, Argonne, IL 60439, USA}
\author{D.~Dutta}
\affiliation{Massachusetts Institute of Technology, Cambridge, MA
02139, USA}
\author{K.~Fissum}
\affiliation{Lund University, S-221 00 Lund, Sweden}
\author{S.~Frullani}
\affiliation{INFN/Sezione Sanita, 00161 Roma, Italy}
\author{H.~Gao}
\affiliation{Massachusetts Institute of Technology, Cambridge, MA
02139, USA}
\affiliation{Duke University, Durham, NC 27708, USA}
\author{R.~Gilman}
\affiliation{Thomas Jefferson National Accelerator Facility, Newport
News, VA 23606, USA}
\affiliation{Rutgers University, New Brunswick, NJ 08903, USA}
\author{C.~Glashausser}
\affiliation{Rutgers University, New Brunswick, NJ 08903, USA}
\author{J.~Gomez}
\affiliation{Thomas Jefferson National Accelerator Facility, Newport
News, VA 23606, USA}
\author{K.~Hafidi}
\affiliation{Argonne National Laboratory, Argonne, IL 60439, USA}
\author{O.~Hansen}
\affiliation{Thomas Jefferson National Accelerator Facility, Newport
News, VA 23606, USA}
\author{D.~W.~Higinbotham}
\affiliation{Massachusetts Institute of Technology, Cambridge, MA
02139, USA}
\author{R.~J.~Holt}
\affiliation{Argonne National Laboratory, Argonne, IL 60439, USA}
\author{C.~W.~de Jager}
\affiliation{Thomas Jefferson National Accelerator Facility, Newport
News, VA 23606, USA}
\author{X.~Jiang}
\affiliation{Rutgers University, New Brunswick, NJ 08903, USA}
\author{E.~Kinney}
\affiliation{University of Colorado, Boulder, CO 80302, USA}
\author{K.~Kramer}
\affiliation{College of William and Mary, Williamsburg, VA 23185, USA}
\author{G.~Kumbartzki}
\affiliation{Rutgers University, New Brunswick, NJ 08903, USA}
\author{J.~LeRose}
\affiliation{Thomas Jefferson National Accelerator Facility, Newport
News, VA 23606, USA}
\author{N.~Liyanage}
\affiliation{Thomas Jefferson National Accelerator Facility, Newport
News, VA 23606, USA}
\author{D.~Mack}
\affiliation{Thomas Jefferson National Accelerator Facility, Newport
News, VA 23606, USA}
\author{P.~Markowitz}
\affiliation{Florida International University, Miami, FL 33199, USA}
\author{K.~McCormick}
\affiliation{Rutgers University, New Brunswick, NJ 08903, USA}
\author{D.~Meekins}
\affiliation{Thomas Jefferson National Accelerator Facility, Newport
News, VA 23606, USA}
\author{Z.-E.~Meziani}
\affiliation{Temple University, Philadelphia, PA 19122, USA}
\author{R.~Michaels}
\affiliation{Thomas Jefferson National Accelerator Facility, Newport
News, VA 23606, USA}
\author{J.~Mitchell}
\affiliation{Thomas Jefferson National Accelerator Facility, Newport
News, VA 23606, USA}
\author{S.~Nanda}
\affiliation{Thomas Jefferson National Accelerator Facility, Newport
News, VA 23606, USA}
\author{D.~Potterveld}
\affiliation{Argonne National Laboratory, Argonne, IL 60439, USA}
\author{R.~Ransome}
\affiliation{Rutgers University, New Brunswick, NJ 08903, USA}
\author{P.~E.~Reimer}
\affiliation{Argonne National Laboratory, Argonne, IL 60439, USA}
\author{B.~Reitz}
\affiliation{Thomas Jefferson National Accelerator Facility, Newport
News, VA 23606, USA}
\author{A.~Saha}
\affiliation{Thomas Jefferson National Accelerator Facility, Newport
News, VA 23606, USA}
\author{E.~C.~Schulte}
\affiliation{Argonne National Laboratory, Argonne, IL 60439, USA}
\affiliation{University of Illinois, Urbana, IL 61801, USA}
\author{J.~Seely}
\affiliation{Massachusetts Institute of Technology, Cambridge, MA
02139, USA}
\author{S.~\v{S}irca}
\affiliation{Massachusetts Institute of Technology, Cambridge, MA
02139, USA}
\author{S.~Strauch}
\affiliation{Rutgers University, New Brunswick, NJ 08903, USA}
\author{V.~Sulkosky}
\affiliation{College of William and Mary, Williamsburg, VA 23185, USA}
\author{B.~Vlahovic}
\affiliation{North Carolina Central University, Durham, NC 2770, USA}
\author{L.~B.~Weinstein}
\affiliation{Old Dominion University, Norfolk, VA 23529, USA}
\author{K.~Wijesooriya}
\affiliation{Argonne National Laboratory, Argonne, IL 60439, USA}
\author{C.~Williamson}
\affiliation{Massachusetts Institute of Technology, Cambridge, MA
02139, USA}
\author{B.~Wojtsekhowski}
\affiliation{Thomas Jefferson National Accelerator Facility, Newport
News, VA 23606, USA}
\author{H.~Xiang}
\affiliation{Massachusetts Institute of Technology, Cambridge, MA
02139, USA}
\author{F.~Xiong}
\affiliation{Massachusetts Institute of Technology, Cambridge, MA
02139, USA}
\author{W.~Xu}
\affiliation{Massachusetts Institute of Technology, Cambridge, MA
02139, USA}
\author{J.~Zeng}
\affiliation{University of Georgia, Athens, GA 30601, USA}
\author{X.~Zheng}
\affiliation{Massachusetts Institute of Technology, Cambridge, MA
02139, USA}

\collaboration{Jefferson Lab Hall A Collaboration}
\noaffiliation
\date{\today}

\begin{abstract}
We have measured the differential cross section for the
$\gamma n \rightarrow \pi^- p$ and $\gamma p \rightarrow \pi^+ n$
reactions
at $\theta_{cm}=90^\circ$ in the photon energy range from 1.1 to 5.5 GeV
at Jefferson Lab (JLab). The data at $E_\gamma \gtrsim 3.3$ GeV exhibit a
global
scaling behavior for both $\pi^-$ and $\pi^+$ photoproduction,
consistent with the constituent counting rule and the existing
$\pi^+$ photoproduction data.
Possible oscillations around the scaling value are suggested by these new
data.
The data show enhancement in the scaled cross section at a center-of-mass
energy near 2.2 GeV.
The cross section ratio of exclusive $\pi^-$ to $\pi^+$ photoproduction
at high energy is consistent with the prediction based
on one-hard-gluon-exchange diagrams.

\end{abstract}


\maketitle
The study of the transition region from nucleon-meson degrees of freedom to 
quark-gluon degrees of freedom in quantum chromodynamics (QCD) for  
exclusive processes is one of 
the most interesting subjects in nuclear physics.
Scaling in the differential cross section $d\sigma /dt$, and hadron helicity conservation have been pursued experimentally as signatures of this transition for years. While global scaling behavior has been observed
in many exclusive processes, no experimental evidence supports 
hadron helicity conservation. Furthermore, the exact nature governing 
the onset of the scaling behavior is not clear.
The relatively large cross section of 
pion photoproduction allows the search
for additional possible signatures: QCD
oscillations and the charged pion cross section ratio. 
In this experiment, three signatures (scaling, QCD oscillations,
charged pion ratio) for the transition are
investigated. 

For an exclusive two-body reaction $AB \rightarrow CD$ at high energy 
and large momentum transfer, the constituent counting rule 
(or the dimensional scaling law) predicts~\cite{brodsky} 
\begin{equation} 
\frac{d\sigma}{dt}(AB \rightarrow CD) \sim {s^{2-n}}f(\theta_{cm}) \ , 
\end{equation}
where $s$ and $t$ are the Mandelstam variables. 
The quantity $n$ is the total number of interacting elementary fields in the reaction,
and $f(\theta_{cm})$ is the angular dependence of the differential cross section.
This rule was originally derived
from dimensional analysis~\cite{brodsky}, and later confirmed within the framework of a
perturbative QCD (pQCD) analysis up to a logarithmic factor of the strong coupling constant $\alpha_s$~\cite{lepage}.
Many exclusive measurements at fixed center-of-mass angles  agree remarkably with this rule~\cite{ppscaling,anderson76,white,disintegration,gilman}.

The applicability of pQCD to exclusive processes 
remains controversial in the few GeV region. The pQCD calculation fails to correctly predict the 
magnitude of the proton magnetic form factor~\cite{isgur}.
Furthermore, the onset of scaling can be sometimes as low as 1 GeV, as shown 
in the deuteron photodisintegration data~\cite{gilman,disintegration}, which is much lower than the scale where pQCD is expected to be valid. 
Moreover, hadron helicity conservation, another consequence of pQCD (This statement is currently under debate~\cite{ralston1}), 
tends not to agree with polarization measurements at JLab in the 
photodisintegration process~\cite{gilman,krishni}, as well 
as in $^1 H(\vec{\gamma},\vec{p})\pi^0$~\cite{krishni2}, $ e d \rightarrow e \vec{d}$~\cite{abbott} and $\vec{e}p \rightarrow e\vec{p}$ elastic scattering~\cite{gep}. 

In addition to the above puzzles, two striking anomalies have been observed 
in $pp$ elastic scattering.
The ratio of 
$(d \sigma/dt)_{\uparrow \uparrow}/(d \sigma/dt)_{\uparrow \downarrow}$ with 
spin normal to the scattering plane can reach 4 at $\theta_{cm}=90^\circ$~\cite{crabb};  
the differential cross section $d\sigma /dt$ oscillates around the scaling value~\cite{hendry}. 
The interference  between the 
short-distance and long-distance (Landshoff) subprocesses due to soft gluon 
radiation~\cite{landshoff} can explain the above spin-spin correlation and oscillatory scaling behavior~\cite{brodsky_lipkin,ralston}. This interference is analogous to the QED effect of
coulomb-nuclear interference observed in charged particle scattering at low energy. Alternatively, the above anomalies in $pp$ scattering
can be interpreted in terms of resonances associated with 
charm production threshold, interfering with a pQCD 
background~\cite{brodsky_de}.

It was previously thought that the oscillatory scaling behavior 
was unique to $pp$ scattering or hadron induced exclusive processes. However, it has 
been suggested that similar oscillations should occur in deuteron 
photodisintegration~\cite{sargsian} and 
pion photoproduction at large center-of-mass angles~\cite{jain00}.  
The recent photodisintegration $d(\gamma,p)n$ data~\cite{gilman,disintegration} showed that the oscillations if present are very weak, and the rapid decrease 
in the cross section with photon energy
(${d\sigma}/{dt}\propto{{s^{-11}}}$) makes it impractical to 
investigate such oscillatory behavior. 
Thus, it is essential to search for oscillations in pion photoproduction, 
which has a much larger cross section at high energy due to a slower decrease in the cross section 
with energy (${d\sigma}/{dt}\propto{{s^{-7}}}$), compared to the 
deuteron photodisintegration. In this Letter, 
we present cross section results for charged pion photoproduction from hydrogen and deuterium at $\theta_{cm}=90^\circ$. This angle was chosen to achieve the highest transverse momentum, which might be the kinematic quantity indicative for the onset of scaling~\cite{disintegration,brodsky_hiller}.

The reaction in which pQCD is expected to manifest itself
at relatively low momentum transfer is the elastic scattering from the
charged pion, since the pion has the simplest valence quark
structure. The study of the charged pion elastic form factor has thus been
pursued vigorously in recent years in both experiment and
theory~\cite{hallcpion,calculations}. The elastic form factor of the charged pion is still dominated by non-perturbative effects at a few 
$\rm (GeV/c)^2$. But these non-perturbative effects cancel out to first order in the differential cross section ratio
${\frac{{\frac{d\sigma}{dt}}(\gamma n \rightarrow \pi^- p)}
{{\frac{d\sigma}{dt}}(\gamma p \rightarrow \pi^+ n)}}$.
Therefore one may expect the $\frac{\pi^-}{\pi^+}$ ratio to give the first
indication of a simple pQCD prediction.  In this letter, we provide data on the
charged pion ratio up to a momentum transfer squared value of 5.0 $\rm (GeV/c)^2$, the highest value ever achieved for this quantity.

Experiment E94-104 was carried out in Hall A~\cite{halla} at the Thomas Jefferson National Accelerator Facility (JLab). The continuous electron beam, at currents around 30 $\mu$A  and energies from 1.1 to 5.6 GeV, impinged on a $6\%$ copper radiator and generated an untagged bremsstrahlung photon beam. 
The production data were taken with the 15 cm cryogenic liquid hydrogen 
(LH2) target for singles $p(\gamma,\pi^+)n$ measurement, or with the liquid deuterium  (LD2) target for coincidence $d(\gamma,\pi^-p)p$ measurement. 
The background was measured by recording data without the production target and with or without the radiator. 
The two High Resolution Spectrometers (HRS) in Hall A, with a momentum resolution of better than $2\times 10^{-4}$ and a horizontal angular resolution of better than 2 mrad, were used to detect the outgoing pions and recoil protons.
Based on two-body kinematics, the incident photon energy was reconstructed from final states, 
i.e. the momentum and angle of the $\pi^+$ in the singles measurement, 
momenta and angles of the $\pi^-$ and $p$ in the coincidence measurement.
Both spectrometers consisted of magnets to focus and bend the charged 
particles (45$^\circ$), vertical drift chambers (VDCs) to record the tracks, 
and scintillator planes (S1/S2) to generate triggers. Two new aerogel 
\v{C}erenkov detectors (A1/A2) in the left spectrometer provided particle identification for positive particles, 
mainly pions and protons, since the time-of-flight technique fails at high momentum. 
The CO$_2$ gas \v{C}erenkov detector and preshower/shower detector in the right spectrometer provided particle identification 
for negative particles, mainly pions and electrons. The average number of detected photoelectrons generated by  a $\beta=1$ particle was approximately 8 
for the aerogel \v{C}erenkov detector A1 ($n=1.015$), approximately 30 for A2 ($n=1.055$) and 
approximately 10 for the gas \v{C}erenkov detector ($n=1.00041$). 


A 100 MeV bin of the reconstructed photon energy spectrum, centered 75 MeV below the beam energy,  was chosen for the data analysis, where the multi-pion contribution is negligible. 
The data after background subtraction, with cuts on trigger type, coincidence timing, PID (particle identification), 
acceptance and photon energy, were compared to the Monte Carlo simulations to extract the raw cross section.
The simulation was done with a JLab Hall A Monte Carlo program, MCEEP~\cite{MCEEP}, with some modifications. The bremsstrahlung photon flux was calculated with an estimated 3\% uncertainty, by using the 
thick-radiator codes written by Meekins~\cite{meekins}, 
based on the formulae of Matthews {\it et al.}~\cite{matthews}. 
The momentum distribution of the neutrons in the deuterium target as well as the binding energy 
was considered in the simulation. 
Different calculations of the momentum distribution were tried and little model dependence 
was found for the cross section ($<1\%$).
 The angular distribution input for the cross sections was fitted from the $\pi^+$ photoproduction data at 4, 5 and 7.5 GeV~\cite{anderson76}, which has not been reproduced by the pQCD calculation~\cite{farrar91}.  
It was used for all the kinematics including the $\pi^-$ photoproduction since the extraction of cross 
 section at $\theta_{cm}=90^\circ$ was insensitive to the angular distribution ($<1\%$).
 The distributions of acceptance, reconstructed momentum 
and photon energy from data were in good agreement with those from 
simulations. 

Several correction factors were applied to deduce the final 
cross section, as shown in Table~\ref{xsection}.  The largest correction, on the order of 20\%,  came from the the nuclear transparency in the deuteron due to final state interactions. The nuclear transparency was obtained for the $d(\gamma,\pi^-p)p$ process based on a Glauber calculation~\cite{haiyan_t}, which has been tested by the measured transparency from the quasi-elastic $d(e,e'p)$ process~\cite{garrow}. 
The correction due to material absorption was applied to compensate for 
the scattering losses in the target and in the spectrometers. The correction for a single 
pion or proton was approximately 6\%, with major losses in the target, 
scintillators and aerogel detectors. The pion decay losses were calculated from the flight distance. As some muons from pion decay may still fall into the acceptance and be mis-identified as pions, an additional correction has to be applied, which was 4$\sim$7\% based on Monte Carlo simulations~\cite{arrington}.
The computer dead time correction was considered run-by-run and was mostly a few percent. 
The detector efficiencies also led to some corrections, mostly less than 1\%.

The total errors were dominated by systematic uncertainties, which were estimated 
to be 8\% in cross section. The point-to-point systematic uncertainty for the three 
kinematics at 3.3, 4.2 and 5.5 GeV is 4\%. The statistical errors were approximately 2\%.
The major systematic uncertainties arose from the calculation of the bremsstrahlung
photon yield, the simulation of the acceptance and the estimation of 
the nuclear transparency, material absorption and pion decay factor, 
 approximately 3\% for each item. Also, there were 2\% 
uncertainties from PID and the energy loss calculations. Other 
systematic uncertainties were less than 1\%.

\begin{figure}[htbp]
\centerline{\includegraphics*[bb=14 157 535 670,width=8cm,height=6cm]{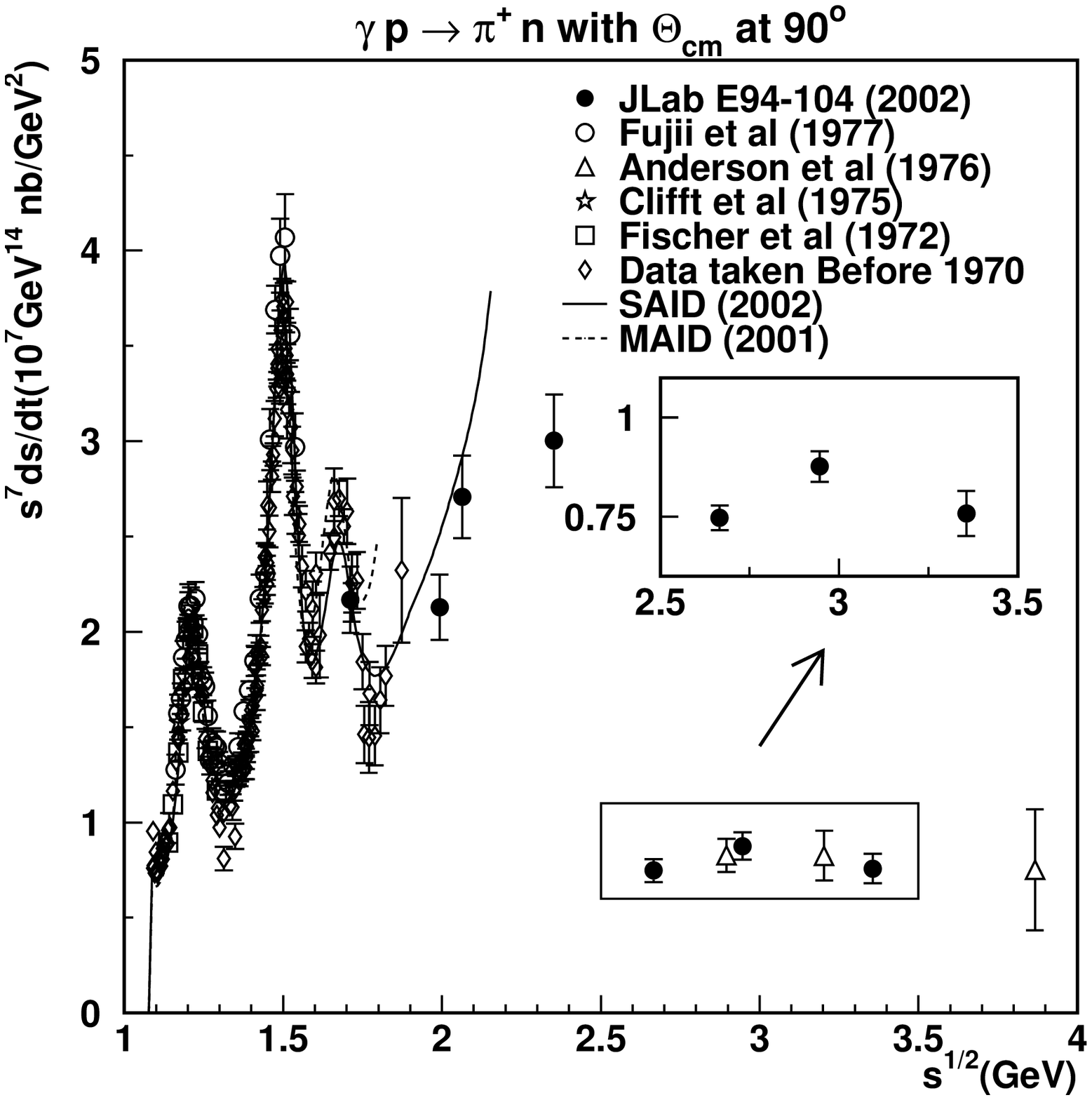}}
\centerline{\includegraphics*[bb=14 157 535 670,width=8cm,height=6cm]{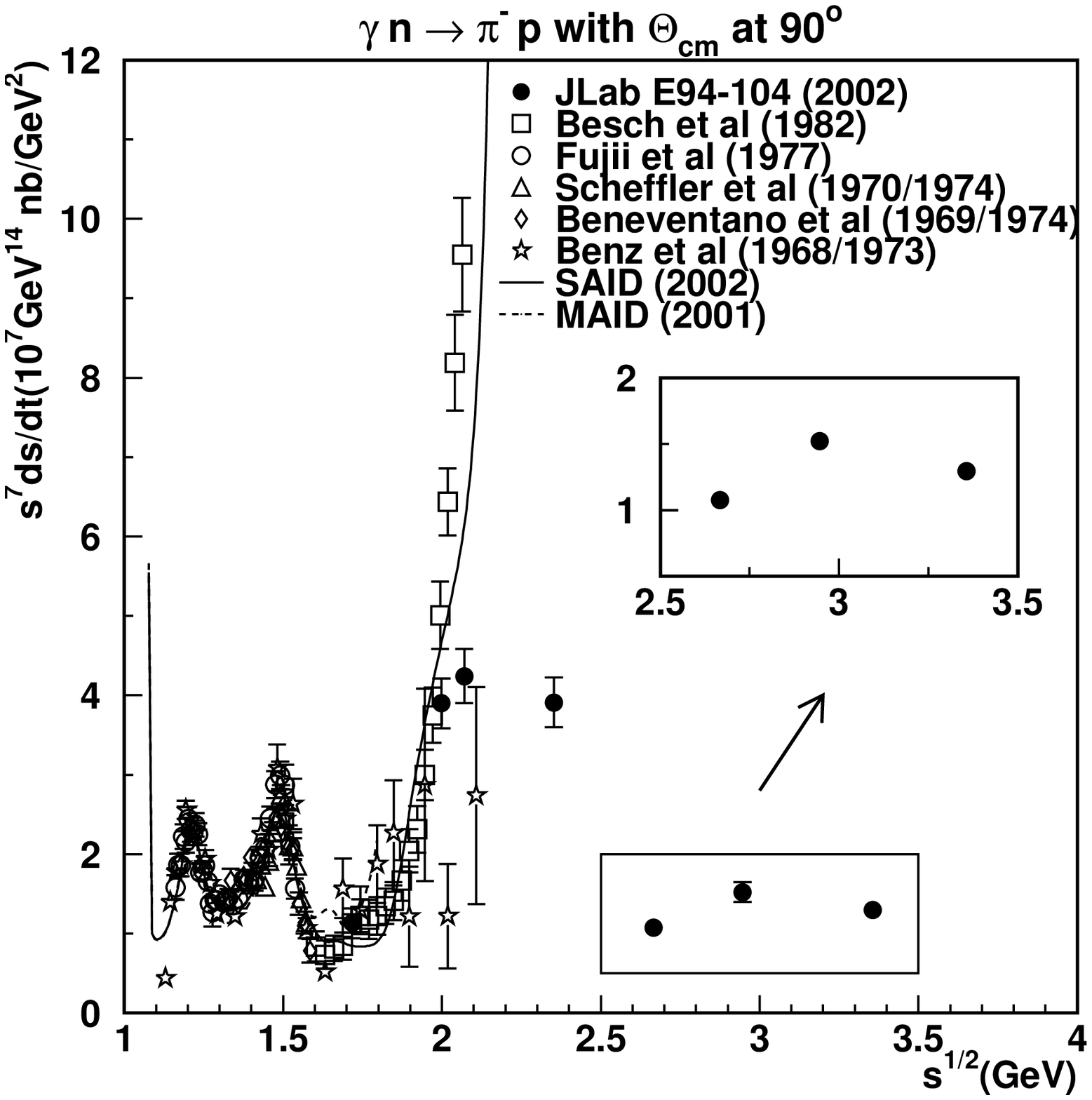}}
\caption[]{The scaled differential cross section $s^{7}{\frac{d\sigma}{dt}}$  versus center-of-mass energy 
for the $\gamma p \rightarrow \pi^+ n$ (upper plot) and $\gamma n \rightarrow \pi^- p$ (lower plot) at 
$\theta_{cm}=90^\circ$. The data from JLab E94-104 are shown as solid circles. The error bars for the new data and Anderson {\it et al.}'s data~\cite{anderson76}, include statistical and systematic uncertainties, except that those in the insets only include point-to-point uncertainties to highlight the possible oscillatory scaling behavior. Other data sets~\cite{world_data1,world_data2} are shown with only statistical errors. The open squares in the lower plot were averaged from data at $\theta_{cm}=85^\circ$ and $95^\circ$~\cite{besch}. The solid line was obtained from the recent partial-wave analysis of single-pion photoproduction data~\cite{said} up to $\rm E_\gamma$=2 GeV, while dash line from the MAID analysis~\cite{maid} up to $\rm E_\gamma$=1.25 GeV.
}
\label{results}
\end{figure}

The upper plot in Fig.~\ref{results} shows the results of the scaled differential cross section
($s^{7}{\frac{d\sigma}{dt}}$) for the $\gamma p \rightarrow \pi^+ n$ process
at $\theta_{cm}=90^\circ$. The new results with fitted value $n=9.0 \pm 0.2$ (see Eq. 1)
 agree with those of Anderson {\it et al.}~\cite{anderson76} and exhibit the 
 scaling behavior predicted by the constituent counting rule with $9$ elementary fields. 
The lowest energy datum in the inset box of Fig.~\ref{results}
corresponds to a center-of-mass energy of approximately 2.7 GeV and photon energy of 3.3 GeV. 
The corresponding transverse momentum is approximately 1.2 GeV/c.

The lower plot in Fig.~\ref{results} shows the results of the scaled differential 
cross section ($s^{7}{\frac{d\sigma}{dt}}$) for the $\gamma n \rightarrow \pi^- p$ process
at $\theta_{cm}=90^\circ$. The new results greatly extend the existing 
measurements and exhibit, for the first time, a global scaling behavior 
at high energy for this reaction with fitted value $n=8.6 \pm 0.2$. 
The scaling behavior in $\pi^-$ production is similar to that 
in  $\pi^+$ production. Furthermore, data in these two channels show 
possible oscillations around the scaling behavior in similar ways 
as suggested by the insets in Fig.~\ref{results}. Note that this 
possible oscillatory
behavior occurs above the known baryon resonance region.
Unfortunately, the coarse and few photon energy settings of this experiment 
do not allow us to claim the observation of oscillations.
Measurements with much finer binning, planned at JLab~\cite{dutta}, 
are essential for the confirmation of such oscillatory scaling behavior. 

The future 12 GeV energy upgrade at JLab will enable us to extend further 
the measurements and study the reactions both below and above the charm production threshold. 
Searching for the resonances around the charm production threshold, 
proposed by Brodsky and de Teramond~\cite{brodsky_de} to explain the 
oscillatory scaling behavior in $pp$ scattering, will help to understand 
the origin of the oscillatory scaling behavior.

\begin{table}[htbp]
\begin{ruledtabular}
\begin{tabular}{|c|cc|}
${\rm E_{\gamma}}$ 
              & ${\frac{d\sigma}{dt}}(\gamma p \rightarrow \pi^+ n)$ 
              & ${\frac{d\sigma}{dt}}(\gamma n \rightarrow \pi^- p)$ 
               \\
(GeV) & (${\rm nb/GeV^2}$) 
              & (${\rm nb/GeV^2}$) 
             \\
\hline
1.106 &(1.16$\pm$0.01$\pm$0.09)$\times 10^4$ & (5.72$\pm$0.03$\pm$0.46)$\times 10^3$  \\ 
1.659  &(1.36$\pm$0.01$\pm$0.11)$\times 10^3$ & (2.39$\pm$0.01$\pm$0.19)$\times 10^3$  \\
1.815  &(1.06$\pm$0.01$\pm$0.08)$\times 10^3$ & (1.58$\pm$0.01$\pm$0.13)$\times 10^3$  \\
2.481  &(1.87$\pm$0.02$\pm$0.15)$\times 10^2$ & (2.43$\pm$0.03$\pm$0.19)$\times 10^2$ \\
3.321  &8.07$\pm$0.09$\pm$0.65 & (1.16$\pm$0.01$\pm$0.09)$\times 10^1$  \\
4.158  &2.34$\pm$0.04$\pm$0.19 & 4.05$\pm$0.08$\pm$0.32 \\
5.536 &0.33$\pm$0.02$\pm$0.03 & 0.56$\pm$0.01$\pm$0.04  \\
\end{tabular}
\end{ruledtabular}
\caption[]{The differential cross section ${\frac{d\sigma}{dt}}$ at $\theta_{cm}=90^\circ$ for $\gamma p \rightarrow \pi^+ n$ and $\gamma n \rightarrow \pi^- p$ reactions followed by the statistical and systematic errors.}
\label{xsection}
\end{table}

Another interesting feature of the data is an apparent enhancement in the 
scaled differential cross section below the scaling region, 
at a center-of-mass energy ranging approximately from 1.8 GeV to 2.5 GeV, 
in both channels of the charged pion photoproduction, as shown in 
Fig.~\ref{results}. This effect was also observed in neutral pion 
photoproduction~\cite{world_data1,world_data2}. 
Without any conclusive statements for the present, some speculations 
can be made. The observed enhancement around 2.2 GeV might 
relate to some unknown baryon resonances, as some of the well known 
baryon resonances ($\Delta$, N$^\star$'s around 1.5 GeV and 1.7 GeV)
are clearly seen in the scaled cross section below 2.2 GeV.
Several baryon resonances are predicted to be in this energy region 
by the constituent quark model~\cite{capstick}, but have not been 
seen experimentally, i.e. the so called 'missing resonances'.
The observed enhancement might be associated with the strangeness 
production threshold.  
When the available energy is near a new flavor threshold, 
all of the quarks have small relative
velocities, allowing resonant behavior in attractive 
channels~\cite{brodsky_de}. Whatever the origin is for the observed enhancement
around 2.2 GeV, it is remarkable that the scaled differential cross section 
drops by a factor of 4 over a rather small center-of-mass energy range 
($\sim$ 0.3 GeV). More experimental work, such as polarization measurements~\cite{dutta2}, together with an extended SAID model might help to understand the exact nature of the observed enhancement.

\begin{figure}[htbp]
\centerline{\includegraphics*[bb=14 157 535 670,width=8cm,height=6cm]{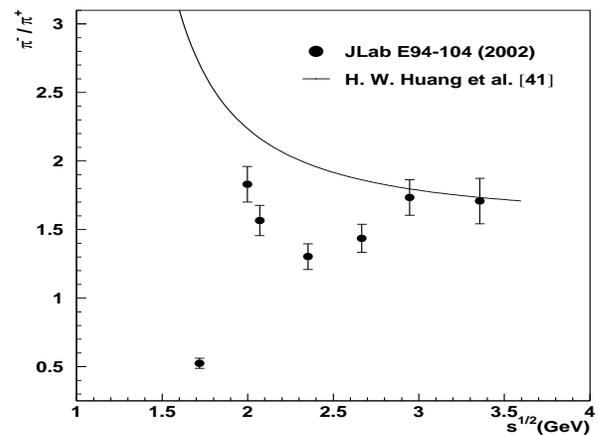}}
\caption[]{The cross section ratio of $\pi^-$ to $\pi^+$ photoproduction  versus center-of-mass energy }
\label{ratio}
\end{figure}

The cross section ratio of $\pi^-$ to $\pi^+$ photoproduction can be 
calculated~\cite{huang} based on one-hard-gluon-exchange diagrams as
\begin{equation}
\frac{d\sigma (\gamma n \rightarrow \pi^- p  )}{d\sigma (\gamma p \rightarrow \pi^+ n  )} \simeq \left( \frac{ue_d+se_u}{ue_u + se_d} \right)^2  \ ,
\end{equation}
where $u$ and $s$ are the Mandelstam variables, and $e_q$ denotes the charge of the quark $q$. The non-perturbative components 
are represented by the form factors which divide out when the ratio is taken.
The calculation is expected to be valid only at high energy. As shown in 
Fig.~\ref{ratio}, the calculation agree with the two data points at the highest energies.

In summary, we have measured the differential cross section $d\sigma /dt$ for 
the photoproduction processes of $\gamma n \rightarrow \pi^- p$ with a deuterium target 
and $\gamma p \rightarrow \pi^+ n$ with a hydrogen target at $\theta_{cm}=90^\circ$ 
with photon energies from 1.1 to 5.5 GeV. The 
data with $E_\gamma \gtrsim 3.3$ GeV exhibit a global 
scaling behavior in both processes, consistent with the constituent counting rule. 
The data with $E_\gamma \gtrsim 3.3$ GeV also suggest a possible 
 oscillatory scaling behavior, the confirmation of which awaits future measurement with finer binning in energy.
Furthermore, the scaled cross section data show an enhancement  
at a center-of-mass energy near 2.2 GeV and the exact nature of 
such a structure requires further investigation.
The data also provide $\pi^-$ to $\pi^+$ cross section ratios, 
consistent with the one-hard-gluon-exchange prediction at high energies.

We acknowledge the outstanding support of JLab Hall A technical staff and
Accelerator Division in accomplishing this experiment.
We thank R.~B.~Wiringa and H.~Arenh$\rm \ddot{o}$vel for calculating the 
momentum distribution of the neutron in the deuteron. 
We thank Z.~Chai for providing the codes to apply R-function cut on acceptance.
We also thank P.~Jain and T.~W.~Donnelly for helpful discussions.
This work was supported in part by the U.~S.~Department of Energy, DOE/EPSCoR,
the U.~S.~National Science Foundation, 
the Ministero dell'Universit\`{a} e della Ricerca
Scientifica e Tecnologica (Murst),
the French Commissariat \`{a} l'\'{E}nergie Atomique,
Centre National de la
Recherche Scientifique (CNRS) and the Italian Istituto Nazionale di Fisica
Nucleare (INFN).
This work was supported by DOE contract DE-AC05-84ER40150
under which the Southeastern Universities Research Association
(SURA) operates the Thomas Jefferson National Accelerator Facility.


\end{document}